\documentclass{PoS}
\usepackage{ amsmath, amsthm, amssymb }
\usepackage{mathtools,slashed}
\title{Production of Higgs bosons with large transverse
momentum}

\ShortTitle{Production of Higgs boson with large transverse
momentum}

\author{\speaker{Kirill Kudashkin}\\
        Karlsruher Institut f\"{u}r Technologie, Institut f\"{u}r Theoretische Teilchenphysik \\
        E-mail: \email{kirill.kudashkin@kit.edu}}
\dedicated{TTP18-028}
\abstract{Recent results on the Higgs plus jet production is reviewed. We have computed the two-loop amplitudes for $ gg \to Hg $, $ q\bar{q} \to Hg $ and $ qg \to Hq $ at the large Higgs transverse momentum. It was long-missing part of NLO corrections to the Higgs plus jet production above the top mass threshold. This results are combined with the real corrections to produce the Higgs boson transverse momentum distribution at NLO. }

\FullConference{Loops and Legs in Quantum Field Theory (LL2018)\\
		29 April 2018 - 04 May 2018\\
		St. Goar, Germany}

\begin{document}

\section{Introduction}
In 2012 the new particle was observed which was later identified with the fundamental scalar boson of the Standard model (SM). Its properties are in perfect agreement with theoretical predictions within SM. Now SM is going through the new period of stress tests where we are increasing the accuracy of our calculations and the precision of our experimental tools such as the Large Hadron Collider (LHC). Both experimental and theoretical physicists are trying to find any signs of discrepancy between the LHC data and SM. 

The top-Yukawa coupling is a potential source of new findings. The possibility to find new physics comes from the fact that the top-Yukawa coupling $ k_t$ is known experimentally to a precision of 50\% from $t\bar{t}H$ production process. It leaves space for a new point-like $ggH$ interaction $ k_g $. To demonstrate this possibility let us look at the inclusive cross-section for $ggH$ in presents of the new point-like interaction, that is, $ \sigma_{gg\to H} \sim \alpha_s^2/\nu^2(k_g + k_t)^2$ where $ k_g, k_t $ are anomalous couplings. In other words if we consider the Higgs production inclusively we will be able to constrain only the sum of anomalous couplings \cite{Grojean:2013nya}. Hence, we must go beyond inclusive searches. Here it should be noted that there was already an attempt to do this experimentally \cite{Sirunyan:2017dgc}. 

The Higgs plus jet production is a way to solve this problem. Indeed, if the Higgs is produced with a additional jet it will result in a complex kinematics which involves both the top quark and the new degree of freedom which is responsible for the new point-like interaction. Assuming the scale of the new physics to be high enough, say $ \lambda_x = 1000\, \text{GeV} $, we will have a region $ 2m_{top} < p_{T,H} < \lambda_x $ where the top loop is resolved and the new interaction can still be considered point-like. Increasing $p_{T,H}$-cut will make the contribution of the top-loop smaller while the point-like interaction will not change \cite{Banfi:2013yoa}. In this way we will be able to break the degeneracy between these two anomalous couplings. 

In this talk we addressed to the problem that there were no sufficiently accurate calculations for the Higgs plus jet production. Indeed, there were only the LO cross-section available in the high $p_{T,H}$ region for 30 years \cite{Ellis:1987xu}. To resolve this problem we have calculated NLO corrections in $ \alpha_s $ to the Higgs boson production at the high $p_{T,H}$ including the top mass effects \cite{Lindert:2018iug}.

\section{NLO corrections to the Higgs plus jet production}
In this section, we consider how the NLO corrections were computed within perturbative QCD. The NLO corrections consist of two parts: a real corrections and a virtual corrections. The real corrections are an one-loop $2\to3$ process. It has been computed analytically \cite{DelDuca:2001eu}. However, we have used an \textit{OpenLoops} implementation for convenience \cite{Cascioli:2011va}.

In general, the virtual correction is difficult to compute. It is a four scale problem: $\{s,t,m_H^2,m_{top}^2 \}$ where $s,t$ are Mandelstam variables, $m_H$ is the Higgs mass and $m_{top}$ is the mass of the top quark. Depending on transverse momentum $p_{T,H}$ of the Higgs, there is a way to simplify calculations. \footnote{For phenomenological reasons, we are assuming that only the top quark is massive. All other quarks are considered massless.} In region where $p^2_{T,H} << m_{top}^2$ one gets the following hierarchy of scales: $ \{s,t,m_H^2\} << m_{top}^2 $. When computing master integrals one can employ the large mass expansion based on the observation above.  

However, this procedure breaks down if $p^2_{T,H} \gg 4m_{top}^2$. Indeed, in this region the hierarchy of scales is the following: $ \ m_H^2 < m_{top}^2 \ll \{s,t\} $. In other words, the large mass expansion is incapable of tacking into account the top loop effects. Thus, one should come up with different approach. Before reviewing this approach from \cite{Kudashkin:2017skd}, let us look at the different works.

The most ambition project is \cite{Bonciani:2016qxi}, where master integrals are being calculated analytically with the full top mass dependence. While planar master integrals are already calculated, non-planar integrals seem to be more complicated. Indeed, the size of intermediate results are so large that it gets harder to handle it. Beyond-polylogarithmic functions also complicate the computations. Namely, one will use symbols to handle this large expressions, but there were no symbols for elliptic functions until recently \cite{Broedel:2017kkb}. \footnote{see the proceedings of H.Frellesvig} More successful way was used in \cite{Jones:2018hbb}. There, two-loop amplitudes were calculated numerically using \textit{SecDec} \cite{Carter:2010hi}. It was capable of getting the full top mass dependence. \footnote{see the proceedings of M. Kerner}

We advocate a different approach. In general, it is based on the differential equations \cite{Kotikov:1990kg}. We will not describe how to get from diagrams to differential equations, since there are many papers where it is written in details \cite{Kudashkin:2017skd}. Here we would like to note an important feature of $p^2_{T,H} >> 4m_{top}^2$ region. It is readily seen that one can construct two small parameters $ m_H^2/4m_{top}^2 \sim 0.13 $ and $ m_{top}^2/s \ll 1 $. Then, it is possible to employ small parameter expansion at the level of differential equations \cite{Melnikov:2016qoc}. Namely, the analysis of the differential equations suggests a particular form of the integrals. We came up with the following ansatz

\begin{equation}
    \mathfrak{I}_i(\kappa,\eta,z,\epsilon)=\sum_{j,k,l,m\in Z,n\in N} c_{i,j,k,l,m,n}(z,\epsilon)\, (m_{top}^2/s)^{j-k\epsilon} (m_H^2/4m_{top}^2)^{l/2-m\epsilon} \, \log^n(m_H^2/4m_{top}^2).
    \label{eq:ansatz1}
\end{equation}
where $\epsilon$ is the dimension regularization parameter, $z = t/s$ and $c_{i,j,k,l,m,n}(z,\epsilon)$ are unknown functions which should still be integrated via differential equations. We emphasis that $ \{i,j,k,l,m,n\} $ indices has to be truncated at some values, that is, it is a finite series. 

Advantages of this method should be noted. First, we are not looking for a canonical form of differential equations. In general, it is not a problem to get a canonical form of differential equations, and at this moment it is better understood then before, but still it requires some effort. Moreover, some of the differential equations for the two-loop master integrals cannot be putted in the canonical form \footnote{Due to elliptic integrals}, but our method will work regardless of the differential equation form. Second, we are effectively reducing the four scale problem to the "one" scale problem. Indeed, after inserting the ansatz into differential equations we will have \textit{algebraic} equations with respect to $ \{m_H^2/4m_{top}^2, m_{top}^2/s\} $ which can be easily solved. At last, the "one" scale problem is well enough understood and can be integrated by known packages \cite{Gituliar:2017vzm}. Finishing the integration will lead us to a point where we have to fix integration constants, that is, the boundary conditions. 

Computing boundary conditions is the last step in the calculations of the virtual corrections. We emphasis that specific physical region is considered and our expansion will only works in this region. It will also affect what kind of limits we can take. For example, $ s \to 0 $ is not possible anymore, since $  m_{top}^2/s \ll 1$ is not satisfied. In general, we cannot take some convenient limits to compute integration constants. However, it is still possible to apply unitarity cuts \cite{Cutkosky:1960sp}. Indeed, the differential equations will have certain type of singularities which don't coincide with singularities of the corresponding diagram, that is, spurious singularities. Hence, we can set corresponding coefficients to be zero when the integral approaches to the singularity. It will result in more algebraic equations and hence allows us to reduce the number of integration constants. It is very powerful method since we fixed roughly 70\% of all 261 constants with this method. The next way can be described as follows: we note that on amplitude level there are \textit{no} singularities in the Higgs mass. Indeed, the Higgs always couples to top quarks. Hence, we must constrain how the ansatz should look like. On the differential equation level we should put to zero all coefficients multiplying any negative powers of the Higgs mass such as $ 1/m_H$ which also results in additional algebraic equations.  We must admit that in few subgraphs it is possible that the external massive line couples to internal massless lines. However, it happened few times and this case can be treated individually. This is just two methods that we applied to compute integration constants. We used different approaches and it is not possible to review them all here. We refer once again to authors' paper \cite{Kudashkin:2017skd}. 

\begin{figure}
    \centering
    \includegraphics[width=\textwidth,keepaspectratio]{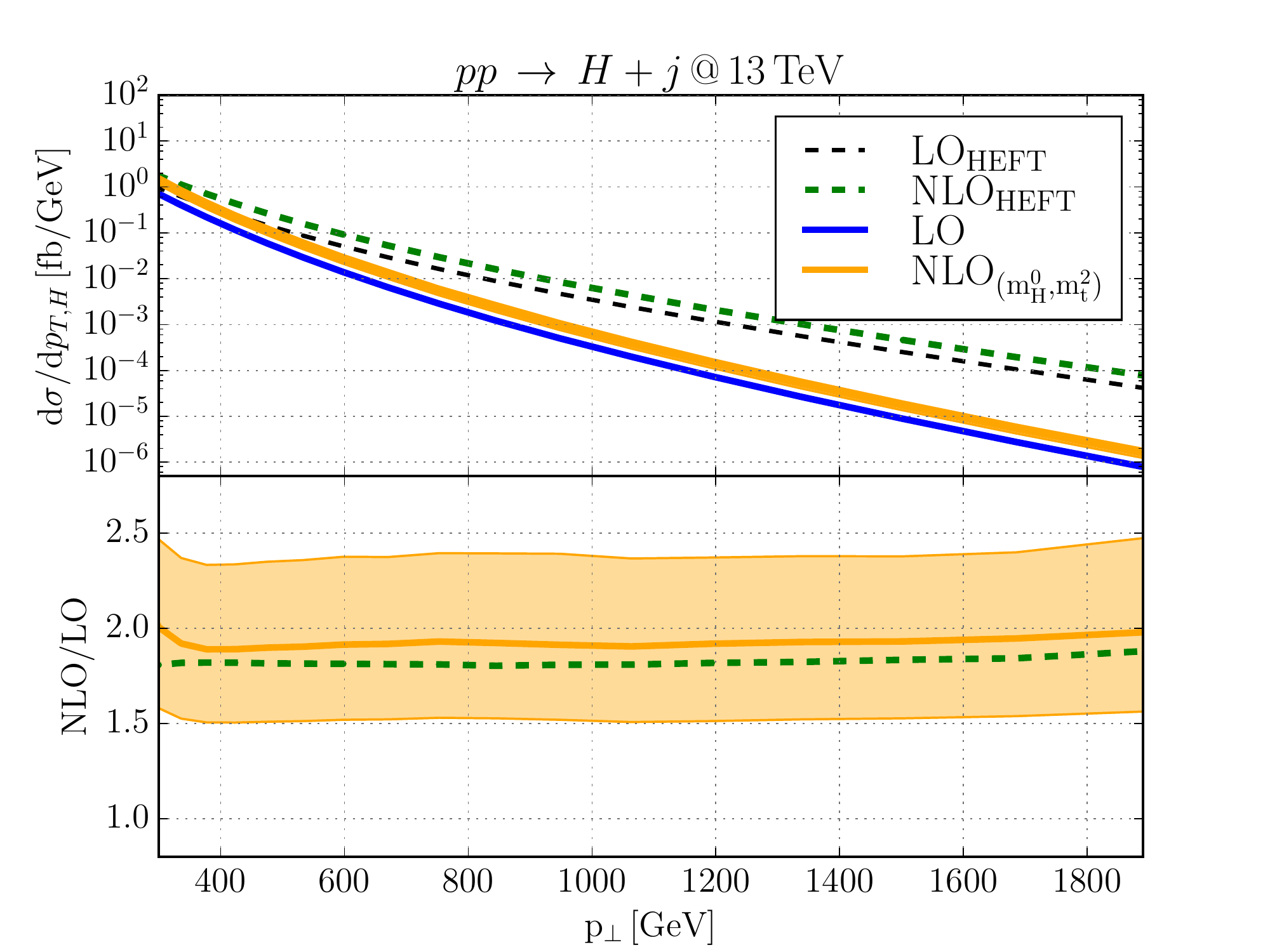}
    \caption{Transverse momentum distribution of the Higgs boson. The upper panel shows absolute predictions at LO and NLO in the small parameter expansion and in the approximation of infinite top mass (HEFT). The lower panel corresponds to the ration between NLO and LO corrections. The bands indicate theoretical uncertainty of the expanded result due to variation of $\mu$.}
    \label{fig:results}
\end{figure}

\section{Results}

Combining real and virtual corrections allows us to present the Higgs boson transverse momentum distribution. We considered proton collisions at $13\, \text{TeV}$. The mass of the Higgs is taken to be $ m_H = 125\, \text{GeV}$. The top mass is taken $ m_{top} = 173.2\, \text{GeV}$. We used five-flavor scheme and considered bottom-quark to be massless parton in the proton. We used NNPDF 3.0 set of parton distributions with provided $ \alpha_s $. The renormalization scale is chosen to be  
\begin{equation}
    \mu_0 = H_T/2,\,  H_T = \sqrt{m_H^2+p^2_{T,H}} + \sum_j p_{T,j}
\end{equation}

We estimate theoretical uncertainties by varying the renormalization and the factorization scales $\mu$ by a factor of two with respect to central value. The final results are shown in a figure \ref{fig:results}.

We note that the ration between NLO and LO correction \footnote{ $K = \sigma_{NLO}/ \sigma_{LO}$ factor } is flat and equal $\sim 1.9$ for $p_{T,H} > 400\, \text{GeV}$. The top mass effects estimated by comparing $K$-factors for the expanded case and HEFT. At $p_{T,H} = 400\, \text{GeV}$ this ration is equal $K_{SM}/K_{HEFT} = 1.04$ and at $p_{T,H} =1000\, \text{GeV}$ this ration is equal $K_{SM}/K_{HEFT} = 1.06$. The theoretical uncertainty is reduced from 40\% at LO to 20\% at NLO. 
\section{Summary and outlook}
Master integrals were computed for virtual corrections to the Higgs plus jet production in the large $p_{T,H}$ region. The virtual corrections are then combined with the existing real corrections in order to produce the Higgs boson transverse momentum distribution. Theoretical uncertainties are estimated to have value 20\%. The main results presented in a figure \ref{fig:results}.
The method of small parameter expansion is proved itself to be useful on this example. It can be applied to different LHC-related phenomenon where the similar hierarchy of scales is present. Our result can be used to constrain anomalous couplings in some BSM scenarios which modify the Higgs sector of SM.
\section*{Acknowledgements} K.K thanks Kirill Melnikov, Christopher Wever, Jonas M. Lindert and Hjalte Frellesvig for useful discussions and for preparations to this talk at the Loops\&Legs 2018. The research of K.K. is supported by the DFG-funded Doctoral School KSETA (Karlsruhe School of Elementary Particle and Astroparticle Physics).

\end{document}